\documentclass[a4paper,11pt]{article}
\usepackage{jinstpub} % for details on the use of the package, please see the JINST-author-manual
\usepackage{lineno}
\title{\boldmath Markov chain Monte Carlo (MCMC) based Likelihood Extraction of Chiral-Odd Compton Form Factors from Deeply Virtual Exclusive Experiments.}

\author[]{Saraswati Pandey, Douglas Q. Adams, and Simonetta Liuti}
\affiliation{Department of Physics, University of Virginia, Charlottesville, VA 22904, USA}

\emailAdd{qbs8mh@virginia.edu}

\abstract{We present a likelihood analysis of observables in deeply virtual exclusive meson production off the proton. The study uses experimental data for cross sections together with beam and target spin asymmetries measured in the unpolarized beams and unpolarized as well as longitudinally polarized target configurations at Jefferson Lab Halls A and B. For a fixed kinematic bin defined by $(Q^2, x_B, t,\epsilon)$, we derive a joint likelihood for the twist-two chiral-odd helicity amplitudes and the corresponding Compton Form Factors (CFFs), that parameterize the underlying QCD scattering amplitude. The joint likelihood analysis of cross-section and Asymmetry data strongly constrains the CFF parameter space beyond the either of the measurements studied alone. Markov chain Monte Carlo (MCMC) methods are employed to explore the full parameter space, extracting the parameter uncertainties, correlations and credible intervals.}


\arxivnumber{2605.18589} 

\begin{document}
\maketitle
\flushbottom

\section{Introduction}
One of the major goals of hadronic physics is to understand and describe the three-dimensional structure of nucleon in terms of its constituent quark and gluon fields. In this regard, experiments such as deep-inelastic experiments have been a great platform for studying the longitudinal momentum distribution of the quarks. 
In recent years, Exclusive scattering processes have been a great source of help in this case by encoding information in terms of Generalized Parton Distributions (GPDs). 

\begin{figure}[ht]
    \centering
    \includegraphics[width=0.35\linewidth]{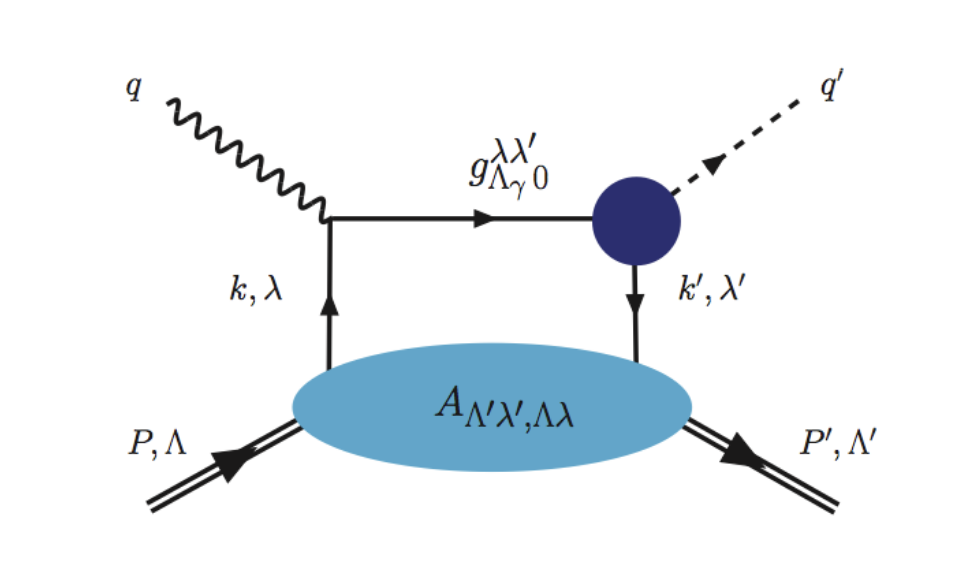}\label{DVMP_process}
    \includegraphics[width=0.4\linewidth]{images/best_kinbin.png}\label{best_bin}
    \caption{(Left) Leading order amplitude for the DVCS/DVMP processes described in the text. Crossed diagrams are not shown in the figure. (Right) The table shows the best chosen set of kinematic bins obtained using Euclidean distance method \cite{enwiki:1325528501} from the experimental data \cite{PhysRevC.83.025201, PhysRevLett.127.152301, HALLB_data}.}
\end{figure}

Deeply Virtual Meson production is an exclusive scattering process analogous to DVCS except that a meson is produced in the final state instead of a real photon and involves helicity flip by the quark. In a basic handbag diagram, the reaction amplitude in a factorized form can be written into two parts: one (top) part describes the basic hard electroproduction process with a parton within the nucleon and the other (down) soft blob, which contains the GPD information-the distribution of partons inside the nucleon resulting from soft processes \cite{PhysRevD.91.114013}. The former is reaction dependent, but the latter is a universal property associated with the nucleon structure for all exclusive reactions. The schematic illustration of the above description is shown in figure \ref{DVMP_process}.

In a QCD factorized scenario, any exclusive process such as Deeply Virtual Meson Production (DVMP) can be expressed as the convolution of hard scattering amplitudes with the specific Generalized Parton Distributions (GPDs). Thus, in total, there are eight GPDs. Out of these, four are parton helicity conserving or chiral-even, $H^q, \Tilde{H}^q,  E^q$, and $\Tilde{E}^q$ and 4 are chiral-odd $H^q _T, \Tilde{H}^q _T,  E^q _T$, and $\Tilde{E}^q _T$ where the parton flips its helicity by one. 
These GPDs depend on four kinematic variables namely: $\rm{x}$, and $\rm{t}$, where $\rm{x}$ is the average parton longitudinal momentum fraction and $\xi$ (skewness) is half of the longitudinal momentum fraction transferred to the struck parton. The skewness is defined in terms of the Bjorken variable $x_B$ as $\xi = x_B/(2 - x_B)$, in which $x_B = Q^2 / 2p.q$, q is the four-momentum of the virtual photon, and $Q^2 = -q^2$. The momentum transfer to the nucleon is $t = (p - p\prime)^2$, where p and $p\prime$ are the initial and final four momenta of the nucleon.

In the forward limit, GPD $H^q_T$ becomes the transversity function  $h^q_1(x)$. Other constraints that exist: Integration in $x$ of the $H^q_T$ gives the elusive tensor charge. Jaffe and Ji \cite{PhysRevLett.67.552} for the first time demonstrated that the first moment of $\mathrm{h_1(x)}$ is related to the nucleon’s tensor charge by:
\begin{equation}
    \int_{-1}^{1} h_1(x) dx = (h_1(x) - \Bar{h}_1(x))dx = \delta \psi
    \label{def_tensor_charge}
\end{equation}
Following equation \ref{def_tensor_charge}, the tensor charge is defined as the net number of transversely-polarized valence quarks in a transversely-polarized nucleon. 

With the recent available data from Jefferson Lab, it is important to determine the size of the various chiral-odd GPDs. The data available so far are not sufficient to fully disentangle the various possible contributions; however, a first essential step is to analyze the available experimental information. At present, the approach we apply for our $\pi^0$ analysis provides a method where the various kinematical dependences can be more readily tested.

We employ Likelihood-Based Bayesian Analysis method to extract the Chiral-Odd Compton Form Factors for Deeply Virtual Exclusive Scattering on the nucleon. A recent study was made to extract Chiral-Even Compton Form Factors for Deeply Virtual Exclusive Scattering on the nucleon. The results were found to produce strong bounds on the CFFs. However, due to lack of sufficient experimental data, strong correlations were observed between the $\mathcal{H}, \mathcal{E}$ and $\mathcal{\Tilde{H}}$ \cite{Adams:2024pxw}.

\section{Likelihood-Based Bayesian Analysis}
%In this work, the statistical framework comprises two steps. First, a joint likelihood function is constructed from the measured cross-sections and spin asymmetries for a chosen kinematic bin. Bayesian inference is performed by combining this likelihood function with the chosen prior distributions for the model parameters, to generate posterior probability distributions for the helicity amplitudes and corresponding Compton Form Factors (CFFs). Second, these posterior distributions are sampled using Markov Chain Monte Carlo (MCMC) techniques to comprehensively explore the allowed parameter space and extract parameter estimates, uncertainties, correlations, and credible intervals.
In this work, the statistical framework consists of two steps. First, a joint likelihood function is constructed from the measured cross sections and spin asymmetries for a given kinematic bin. Using Bayesian statistics, the likelihood is combined with prior information, or prior probabilty distribution, to define the posterior distribution for the helicity amplitudes and the corresponding Compton Form Factors (CFFs). In the second step, the posterior distribution is explored using Markov Chain Monte Carlo (MCMC) sampling, allowing the complete space of statistically admissible solutions to be mapped {\it rather than restricting the analysis to a single best-fit point and its local neighborhood}. The resulting posterior samples provide a comprehensive characterization of the inference problem, including parameter estimates, Bayesian credible intervals, correlations, non-Gaussian features, and potential degeneracies, thereby revealing the full information content of the experimental data.

\subsection{Experimental Data Used in this Analysis}
For this analysis, we have chosen Deep Exclusive Electroproduction of $\pi ^0$ at High $Q^2$ in the Quark Valence Regime from the Jefferson HALL A collaboration with the kinematic parameters $[E, x_{Bj}, Q^2, t, \epsilon]$. Another cross-section data that we have used is the Exclusive neutral pion electroproduction in the deeply virtual regime \cite{PhysRevC.83.025201} again from the Jefferson HALL A collaboration.

In a concise way, we have made an attempt to make the best use of the experimental data in order to study the Compton form factors. If we look at the experimental data available so far from the HALL A and HALL B, we find that for a given kinematic bin $[E, x_{Bj}, Q^2, t, \epsilon]$, there is only one data point for each of the structure functions and asymmetries. Therefore, we optimize the available data by calculating the euclidean distance between each of the rows in the available data for the cross-sections and the asymmetries ($\mathrm{A_{UL}, A_{LL}}$) \cite{enwiki:1325528501}. The pair for which this distance is shortest, are the two most closest kinematic bins. These bins are mentioned in the table on the right side of Figure \ref{best_bin}. The average of these two bins is used for the analysis. Of course the analysis can be performed for each set of kinematic bin but we defer this for future work. 

\subsection{Likelihood Analysis}
Let us consider a random sample $\mathrm{X_1, X_2,\ldots,X_n}$ whose assumed probability distribution is dependent on some unknown parameter $\theta$. In such a scenario, our primary objective is to construct a point estimator u($\mathrm{X_1, X_2,...,X_n}$), such that  u($\mathrm{x_1, x_2,\ldots,x_n}$) is a good point estimate of $\theta$ where $\mathrm{x_1, x_2,\ldots,x_n}$ are the observed values (experimental data) of the random sample. In other words, a good estimate of the unknown parameter $\theta$ would be the value of $\theta$ that maximizes the probability or the likelihood of getting the data we observed. This is in general the idea behind maximum likelihood estimation. A deeper understanding of this can be seen in reference \cite{hogg2010dataanalysisrecipesfitting}. In practice, this is implemented as: 

\begin{equation}
    \mathcal{L}(\theta) = P(X_1 = x_1, X_2 = x_2, \dots, X_n = x_n)
                = f(x_1; \theta) \cdot f(x_2; \theta) \cdots f(x_n; \theta)
                = \prod_{i=1}^{n} f(x_i; \theta)
\end{equation}
where $f(x_i;\theta)$ is the probability density function of the model.

A detailed explanation of the method is mentioned in our recent work \cite{Adams:2024pxw}. For a fixed set of kinematic parameters $[E, x_{Bj}, Q^2, t]$, the total cross-section for exclusive $\pi^0$ electroproduction can be expressed in terms of structure functions arising from the contributions of different helicity amplitudes asscociated with the initial and final particle helicities. The total cross-section is written as:

\begin{eqnarray}
\label{cross-section} 
\hspace{-4ex}
\frac{d^4\sigma}{d x_{Bj} dy d\phi d|t| } & = &
 \Gamma\left\{ \Big[  F_{UU,T} + \epsilon F_{UU,L}+ \epsilon \cos 2\phi F_{UU}^{\cos 2 \phi} 
\right.  
+ \sqrt{\epsilon(\epsilon+1)} \cos \phi F_{UU}^{\cos \phi} \nonumber \\
& + & h \sqrt{2\epsilon(1-\epsilon)} \sin \phi F_{LU}^{\sin \phi} \Big] + S_{\parallel} \Big[ \sqrt{\epsilon(\epsilon+ 1)}\sin \phi  F_{UL}^{\sin \phi} +  \epsilon \sin 2 \phi F_{UL}^{\sin 2\phi} \nonumber \\
& + &  (h)  \left( \sqrt{1-\epsilon^2} F_{LL} + \sqrt{ \epsilon(1- \epsilon)} \cos \phi F_{LL}^{\cos \phi}  \right) \Big] \nonumber \\
& - & (S_\perp) \Big[ \sin(\phi - \phi_S) \left( F_{UT,T}^{\sin(\phi - \phi_S) } + \epsilon F_{UT,L}^{\sin(\phi - \phi_S) }\right) + \frac{\epsilon}{2} \Big( \sin(\phi + \phi_S) F_{UT}^{\sin(\phi + \phi_S)} \nonumber \\
& + & \sin(3\phi-\phi_S)F_{UT}^{\sin(3\phi - \phi_S)} \Big) + \sqrt{\epsilon(1+\epsilon)} \left( \sin\phi_S F_{UT}^{\sin\phi_S} + \sin(2\phi-\phi_S) F_{UT}^{\sin(2\phi-\phi_S)} \right) \Big] \nonumber \\
& + & (S_\perp h)\Big[ \sqrt{1-\epsilon^2} \cos(\phi-\phi_S) F_{LT}^{\cos(\phi-\phi_S)}  \nonumber \\
& + & \sqrt{\epsilon (1-\epsilon)} \Big( \cos\phi_S F_{LT}^{\cos\phi_S} + \cos(2\phi-\phi_S) F_{LT}^{\cos(2\phi -\phi_S)} \Big) \Big] \Big\}
\end{eqnarray}
where, the longitudinal spin is $S_{||} \equiv \Lambda$,  while $\mid {\vec S}_T \mid \equiv \Lambda_T$, $S_T$ being the transverse proton spin at an angle $\phi_S$ with the lepton plane.

In this work, we use the above cross-section model presented by Ref.\cite{PhysRevD.91.114013}. %\textcolor{blue}
{Assuming the model provides adequate description of the underlying physics, the calculated structure functions (and hence the cross-section) still differ from the observed ones. Following the experimental analysis, the standard deviation for each data point is taken as the total experimental uncertainty, with statistical and systematic uncertainties added in quadrature. The resulting systematic uncertainties average approximately $\mathrm{4.5\%}$ and smalller than the statistical uncertainties for most kinematic bins. The difference between the model predictions and the observed data are assumed to follow independent Gaussian distributions with these standard deviations \cite{PhysRevC.83.025201, HALLB_data}. Appropriately, we use a univariate Gaussian likelihood for each data point to evaluate the probability of the observed data. This assumes statistically independent measurements and neglects correlations among experimental uncertainties. For our model parameters, we use improper uniform prior over the interval $(-100, 100)$, thereby normalizing the posterior distribution (proportional to the likelihood) using Bayes’ theorem. The full likelihood is given by taking the product over all data points. These posterior distributions
are subsequently sampled using the MCMC algorithm which take multidimensional probability density functions and generate a set of representative samples which have the underlying distribution, consequently, exploring the high-dimensional parameter space and extract parameter estimates, uncertainties, and correlations.}

%\textcolor{blue}
{The posterior distributions are sampled using the affine-invariant ensemble MCMC sampler implemented in the \texttt{emcee} package. A more detailed explanation of the \texttt{emcee} ensemble sampling algorithm which very close to our problem type is provided in \cite{Adams:2024pxw}. For each kinematic bin, an ensemble of $N_w = 144$ and $64$ walkers (for helicity amplitudes (HA) and CFFs, respectively) is initialized uniformly within the bounded parameter domain defined by the prior ranges. No explicit burn-in period was discarded, and duplicate samples were removed prior to the subsequent analysis. The estimated integrated autocorrelation times for the parameters ranged from approximately $\mathrm{(3.1-3.6) \times 10^3}$ steps. The convergence is monitored by inspecting trace of the walkers and stability of posterior summaries across independent runs. The mean acceptance fraction of the ensemble was approximately $\mathrm{0.06}$.}

\section{Results and Discussions}
\label{results}

In a QCD factorized scenario, the total cross-section can be expressed in terms of structure functions arising from the various combinations of helicity amplitudes. However, the important question here is what is the observable that can be even extracted from experiments if we weren't in a zone of QCD factorization - that would be the helicity amplitudes. Therefore, in our analysis we try to test factorization from the experimental viewpoint. Using the above described method of analysis, we study the chiral-odd Compton form factors and the more fundamentally, the helicity amplitudes corresponding to these CFFs. The corresponding results for these CFFs are as follows:

\begin{figure}
    \centering
    \includegraphics[width=1.0\linewidth]{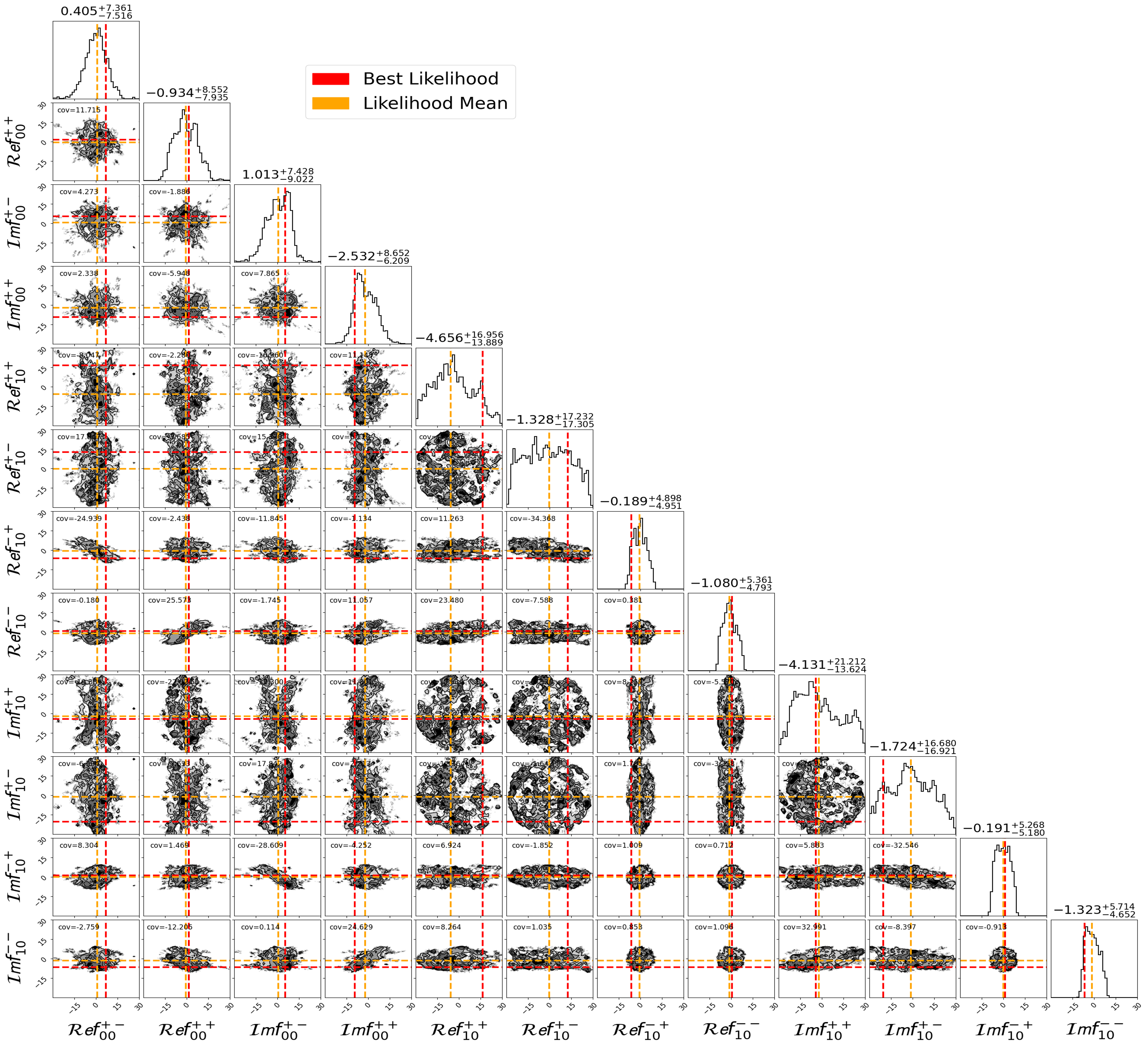}
    \caption{Corner plot for the chiral-odd helicity amplitudes using the last 10000 MCMC samples obtained from the joint analysis of cross-section and Asymmetry data experimental data \cite{HALLB_data, PhysRevLett.127.152301, PhysRevC.83.025201}. Black: MCMC posterior samples, Red: Maximum-likelihood sample, Orange: likelihood-weighted mean.}
    \label{fig_HA}
\end{figure}

The helicity amplitudes are presented in Figure \ref{fig_HA} in the form of corner plots from the unique MCMC posterior samples. The values reported above the 1D-histograms correspond to the posterior median and the central 68\% posterior credible interval, defined by the $\mathrm{16^{th}, 50^{th}}$, and $\mathrm{84^{th}}$ percentiles of the posterior samples for each parameter. The marker ``Best Likelihood'' in the figure denotes that the sampled parameter set maximizes the likelihood among the displayed MCMC samples. The ``Likelihood Mean'' marker denotes the likelihood-weighted average of the displayed MCMC samples. This is solely for visualization purpose and not used as the quoted parameter estimate.

\begin{figure}
    \centering
    \includegraphics[width=1.0\linewidth]{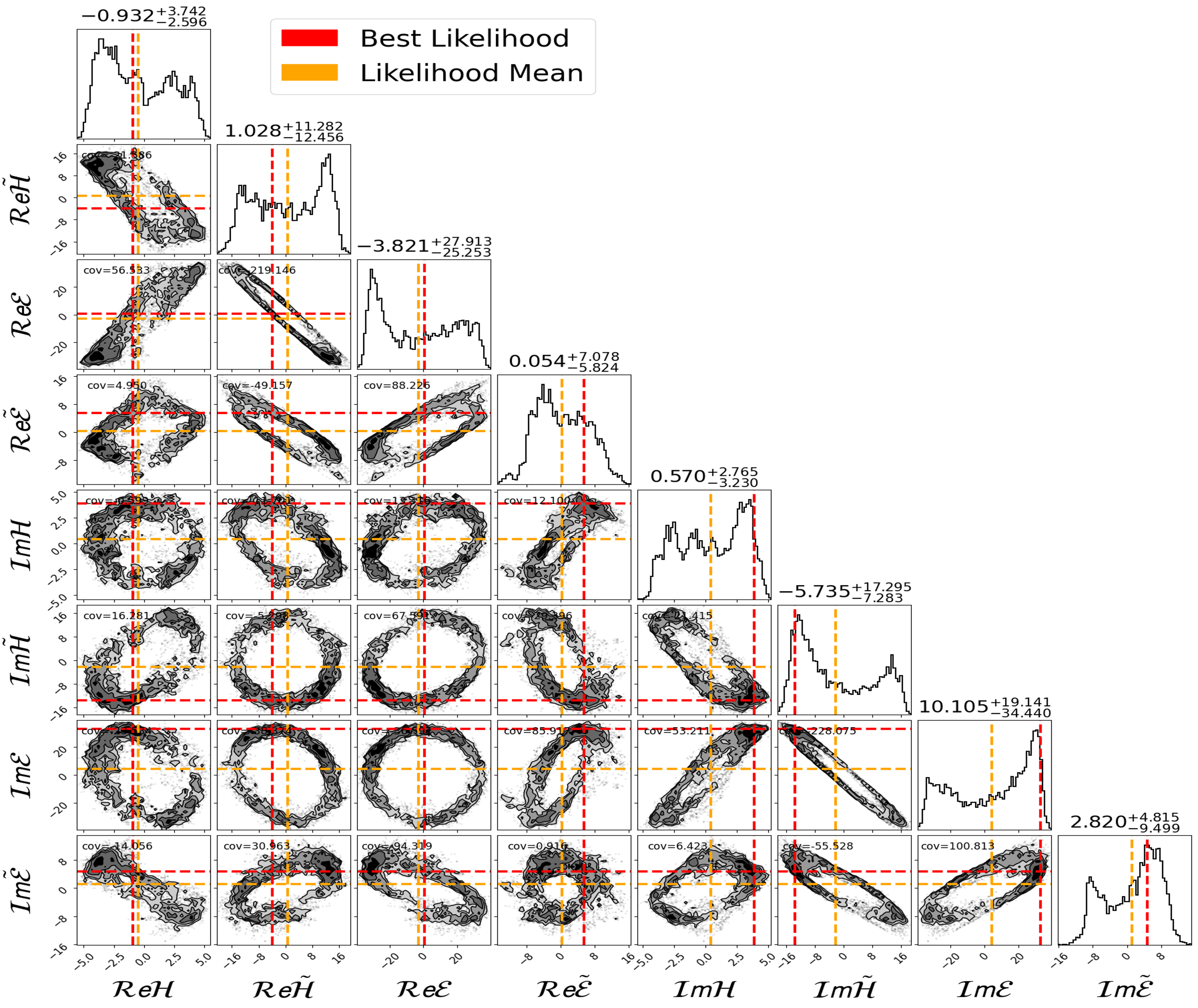}
    \caption{Corner plot for the corresponding chiral-odd Compton form factors using the last 10000 MCMC samples obtained from the joint analysis of cross-section and Asymmetry data \cite{HALLB_data, PhysRevLett.127.152301, PhysRevC.83.025201}. Black: MCMC posterior samples, Red: Maximum-likelihood sample, Orange: likelihood-weighted mean.}
    \label{fig_CFFs}
\end{figure}

\begin{figure}
    \centering
    \includegraphics[width=1.0\linewidth]{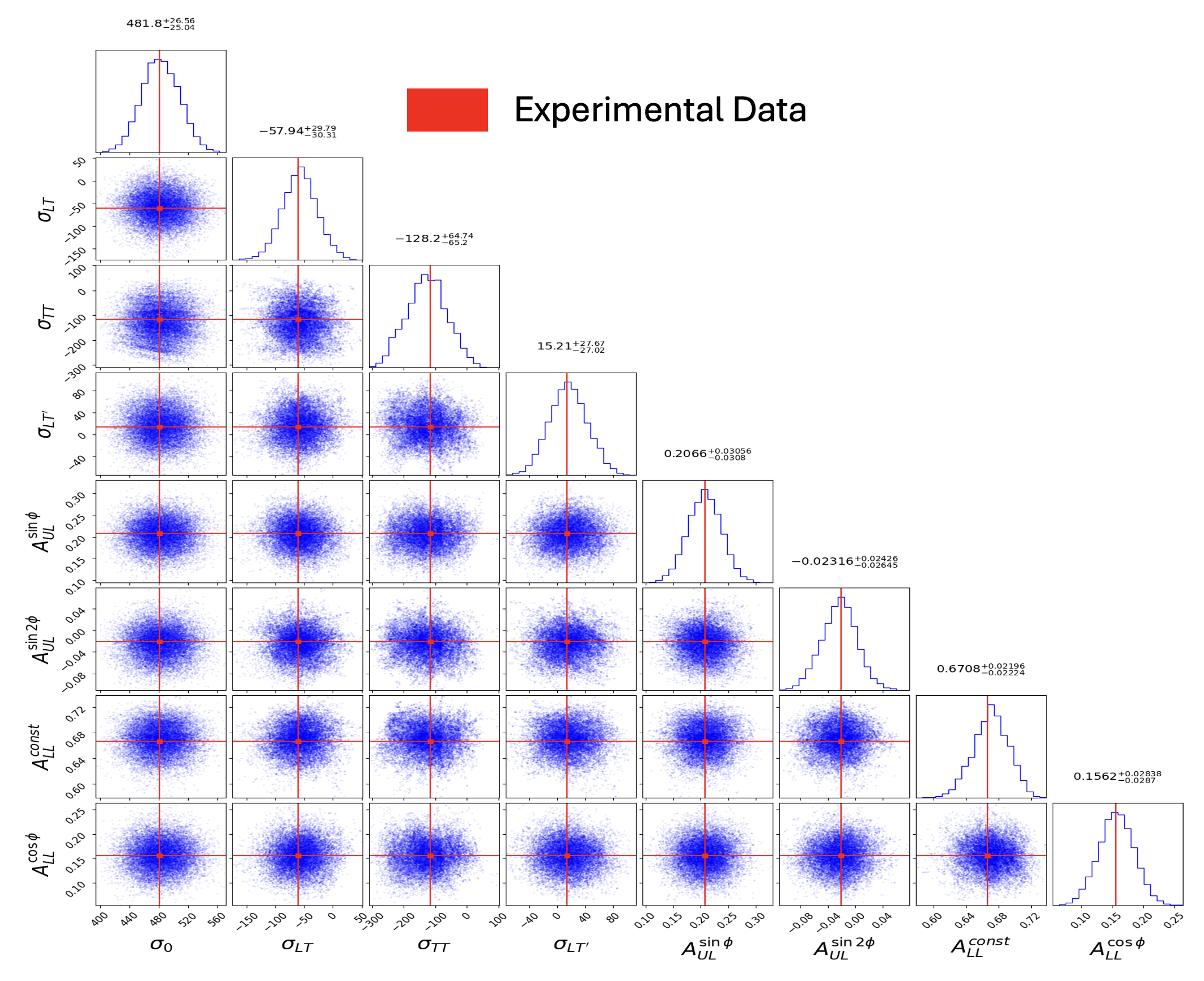}
    \caption{Posterior-predictive comparison of the inferred observables obtained from the joint analysis of cross-section and Asymmetry data with the experimental observations \cite{HALLB_data, PhysRevLett.127.152301, PhysRevC.83.025201}. Blue: last 10000 MCMC samples for the reconstructed cross-sections and Asymmetry analogous to experimental data. Red: Experimental data at the kinematic bin $Q^2 = 2.75$ $\mathrm{GeV^2}$, $x_{Bj} = 0.44, |t| = 0.495$ $\mathrm{GeV^2}$, $\epsilon = 0.545]$.}
    \label{fig_consistency}
\end{figure}

The corner plots show that the helicity amplitudes can be constrained by the available experimental data so far, while also revealing the existing correlations among the parameters. The corresponding Chiral-Odd Compton Form Factors are shown in Figure \ref{fig_CFFs}. The extracted CFFs exihibit strong correlations (hyperspheric behavior), meaning, the experimental data constrain particular combinations of the CFFs rather than each individual CFF independently. This behavior is similar to the theoretical structure of unpolarised cross-section ($F_{UU,T}$) in an exclusive $\pi^0$ electroproduction \cite{PhysRevD.91.114013}, also aligning with the predominant $F_{UU,T}$ in the experimental data. For several parameters, the separation between the Likelihood Mean and Best Likelihood is attributed to the structure of the likelihood surface. The extended degeneracy (in the multidimensional parameter space) in the likelihood leads to broad and strongly correlated posterior distributions. Therefore, the maximum likelihood region is relatively narrower in the parameter space while the posterior mass exists all over the space. The strong degeneracy and thereby correlations indicate that there is significant scope of improvement in the analysis in the sense that we need experimental data to disentangle the helicity amplitudes or CFFs. In this regard, more data from polarized experiments (polarized cross-sections or structure functions) will significantly improve the outcomes. Here, we intentionally use uniform priors but more informative priors can constrain the parameters and improve the analysis. The predictive power of the analysis is illustrated in Fig.~\ref{fig_consistency}, where the final 10000 MCMC samples are compared with the experimental measurements. The  Monte-Carlo samples show a good agreement with the experimental data indicating that the extracted parameter distributions reproduce the experimental data within the quoted experimental uncertainties.

Although the mean acceptance fraction is typically lower than the simple posteriors, it is obvious to expect this value for a high-dimensional (12 or 8 parameters). This indicates that the likelihood surface is strongly correlated, where the proposed moves of the walkers are rejected down due to narrow and curved degeneracy. The autocorrelation time values are stable across parameters. This behavior is attributed to strong correlations in the posterior. This is one of the crucial issues which arise in global fits, as shown, for instance, in the strong correlations appearing in the CFFs extraction from deeply virtual exclusive scattering have  \cite{Adams:2024pxw}. In our analysis, we expect that the marginalized posterior distributions and estimated parameters shall remain stable even under larger sample count indicating robustness of the results.

\section{Conclusions/Summary}
\label{summary}
We presented a new analysis of deeply virtual exclusive $\pi^0$ electroproduction off a proton target, $ep \rightarrow e' \pi^0 p'$, using a model for likelihood-based inference in high dimensional data settings. We obtained joint (statistically) covariant results for Chiral-Odd Compton form factors using a twist-two cross section model for the unpolarized process for experimental data with kinematic dependences available from Jefferson Lab.
\acknowledgments
This work was completed under a grant by the EXCLAIM collaboration under the Department of Energy grant DE-SC0024644. We are thankful to the collaboration for fruitful discussions.


\begin{thebibliography}{99}

\bibitem[]{PhysRevD.91.114013}
Goldstein, Gary R. et al.
\emph{Flexible parametrization of Generalized Parton Distributions: The chiral-odd sector},
\emph{Phys. Rev. D} {\bf 91} {\bf 11} (2015) 114013.

\bibitem{enwiki:1325528501}
Wikipedia contributors, ``Euclidean distance,'' 
\emph{Wikipedia, The Free Encyclopedia}, 2025. 
\emph{[Online]. Available: \texttt{https://en.wikipedia.org/w/index.php?, title=Euclidean\_distance\&oldid=1325528501.}} [Accessed: 31-Mar-2026].

\bibitem[]{HALLB_data}
A. Kim, H. Avakian, V. Burkert, K. Joo et al.
CLAS collaboration
\emph{Target and double spin asymmetries of deeply virtual $\pi^0$ production with a longitudinally polarized proton target and CLAS},
\emph{Phys. Lett. B} {\bf 768} (2017) 168-173.

\bibitem[]{PhysRevLett.127.152301}
Dlamini, M., Karki, B., Ali, S. F., et al.
Jefferson Lab Hall A Collaboration
\emph{Deep Exclusive Electroproduction of ${\ensuremath{\pi}}^{0}$ at High ${Q}^{2}$ in the Quark Valence Regime}
\emph{Phys. Rev. Lett.} {\bf 127}, {\bf 15} (2021) 152201.

\bibitem[]{PhysRevC.83.025201}
Fuchey, E., Camsonne, A., Mu\~noz Camacho, C., Mazouz, M., et al.
Jefferson Lab Hall A Collaboration
\emph{Exclusive neutral pion electroproduction in the deeply virtual regime}
\emph{Phys. Rev. C} {\bf 83}, {\bf 2} (2011) 025201.

\bibitem[]{PhysRevLett.67.552}
Jaffe, R. L. and Ji, Xiangdong
\emph{Chiral-odd parton distributions and polarized Drell-Yan process},
\emph{Phys. Rev. Lett.} {\bf 67}, {\bf 5}(1991) 552--555.

\bibitem[]{physletb.2024.138459}
A. Kim, S. Diehl, K. Joo, V. Kubarovsky et al.
CLAS collaboration
\emph{Beam spin Asymmetry measurements of deeply virtual $\pi^0$ production with CLAS12},
\emph{Physics Letters B} {\bf 849} (2024) 138459.

\bibitem[]{Adams:2024pxw}
Douglas Q. Adams, Joshua Bautista, Marija Cuic, et al.
\title{Likelihood and Correlation Analysis of Compton Form Factors for Deeply Virtual Exclusive Scattering on the Nucleon},
\emph{arxiv:2410.23469}.

\bibitem{hogg2010dataanalysisrecipesfitting}
D.~W.~Hogg, J.~Bovy, and D.~Lang,
\emph{Data analysis recipes: Fitting a model to data},
\emph{arXiv preprint} arXiv:1008.4686 [astro-ph.IM] (2010).

\end{thebibliography}
\end{document}